\documentclass[aps,pra,showpacs,amsmath,twocolumn,superscriptaddress]{revtex4}
\usepackage{amsmath,amssymb,graphicx}
\usepackage[final]{hyperref}
\usepackage{booktabs}

\hypersetup{
	colorlinks=true,       
	linkcolor=blue,          
	citecolor=blue,        
	filecolor=magenta,      
	urlcolor=blue
	}

\begin{document}

\title{Quantum-Enhanced Two-Photon Spectroscopy Using Two-mode Squeezed Light}

\author{Nikunjkumar Prajapati}
\author{Ziqi Niu}
\author{Irina Novikova}
\affiliation{Department of Physics, William \& Mary, Williamsburg, VA 23187, USA}


 \date{\today}



\begin{abstract}
We investigate the prospects of using two-mode intensity squeezed twin-beams, generated in Rb vapor, to improve the sensitivity of spectroscopic measurements by engaging two-photon Raman transitions. As a proof of principle demonstration, we demonstrated the quantum-enhanced measurements of the Rb $5D_{3/2}$ hyperfine structure with reduced requirements for the Raman pump laser power and Rb vapor number density.
\end{abstract}



\maketitle

Measurements of the  energy level structure of atoms and molecules relied on optical absorption and transmission resonances for over a century, and they are still one of the most common spectroscopical tools~\cite{modern_spectroscopy}. 
Typically, a broadband or broadly-tunable light source is used to determine the characteristics of various optical transitions through their absorption spectrum. In some applications, however, the amount of light allowed to interrogate the system is limited due to potential temporary or permanent photo-damage~\cite{Rasmussen2441}, or the available light power is a commodity, like in space exploration or miniaturization ~\cite{problems_in_space}.
Under such constraints, the measurement sensitivity is ultimately limited by the best attainable signal-to-noise ratio scaling as the square root of the detected photon number~\cite{marino_oam_2008,quant_sense:pooser:2018}.
The quantum-enhanced detection methods that take advantage of the non-classical photon correlations, such as squeezing, can boost measurement sensitivity beyond this classical shot noise limit, and thus offer a promising pathway towards performance improvements in the situations where the further increases in the optical power of the probe has negative implicaitons~\cite{GW_detect_with_squeezed:Chau:2014,TAYLOR20161}. For example, intensity squeezed light has been used to overcome quantum shot noise limit in measurements of living systems~\cite{bowen2013NP,Rasmussen2441,Jermyn_2016}, reducing overall power requirements. However, many squeezed light sources are only available at limited ranges of wavelength and usually are not easily tunable, which limits their applicability for spectroscopy. 

\begin{figure*}[htbp!]
	\centering
	\includegraphics[width=0.9\textwidth]{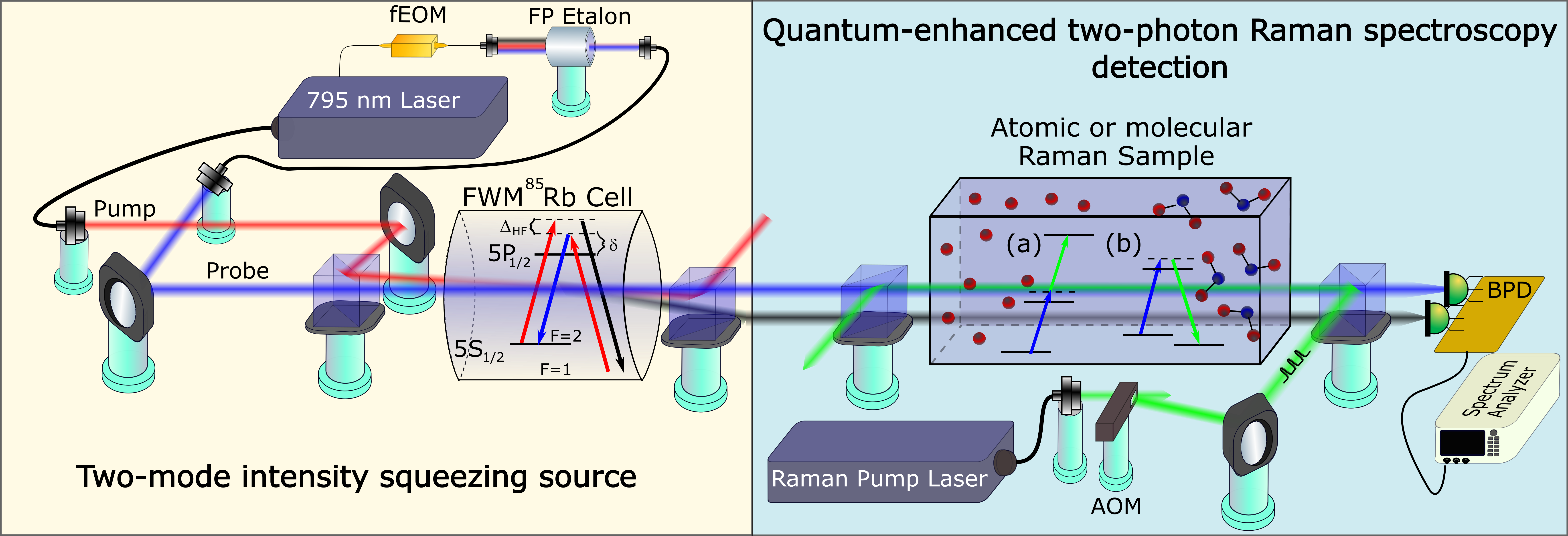}
	\caption{Conceptual experimental setup of the quantum-enhanced spectroscopy experiments using two-mode intensity squeezing generated in ${}^{85}$Rb vapor via four-wave mixing (FWM). An additional strong Raman pump provides access to the levels of interest in a sample either via ladder (a) or $\Lambda$ (b) two-photon Raman transitions, and the resulting absorption resonances are detected in differential intensity of the squeezed twin beams. See text for the abbreviations}
	\label{fig:setup_raman}
\end{figure*}

In this paper, we propose a method to circumvent the problem of limited tunability of squeezed light sources by measuring two-photon Raman resonances, formed by a fixed-frequency quantum probe and tunable classical pump. In these experiments we use intensity-squeezed twin beams, generated by four-wave mixing (FWM) in warm Rb vapors~\cite{lettPRA08,lettSci08,quant_sense:pooser:2018,LettOE2019}. Such twin-beams share quantum fluctuations, and thus their intensity difference allows measurements of small absorption signals below the shot noise limit. Previous experiments demonstrated $\approx 9$~dB of differential intensity noise reduction below the shot noise limit~\cite{lettPRA08,9dbsqueeze}, and have been applied for various quantum-enhanced sensing applications~\cite{quant_sense:pooser:2018,PhysRevA.95.063843Lett,Prajapati:19}. However, since the efficient generation of twin beams is confined to a small frequency range near the Rb atomic resonances~\cite{PooserOE09,Jing_2017,Prajapati19OL}, such quantum light was not used for spectroscopic applications.

The concept of the proposed quantum-enhanced spectroscopy measurements is shown in Fig.~\ref{fig:setup_raman}.
By utilizing two-photon Raman resonances, we can extend the potential quantum advantage of squeezed light into spectral regions for which no such sources are available, using an additional tunable pump field to induce two-photon absorption (or transmission) resonances~\cite{FleischhauerRevModPhys05,PhysRevA.55.2459}. 
By tuning the pump frequency and monitoring the resulting change in the quantum probe transmission we can map the positions of the atomic levels. 
By using a two-mode intensity-squeezed twin beams, we can increase the sensitivity of the absorption measurements, thus softening the laser power requirements and/or concentration of the probed material.
Such coupling can be performed in either a ladder or $\Lambda$ configurations, shown in Fig.~\ref{fig:setup_raman} (a) and (b), respectively. 
The ladder configuration is particularly well-suited for studies of highly excited states of atoms, such as Rydberg states~\cite{8878963}, or as a multi-photon spectroscopy tool~\cite{Michael2019,PhysRevLett.105.173602}. 
The $\Lambda$-configuration is attractive for spectroscopic measurements of the vibrational states of molecules~\cite{doi:10.1119/1.2173278}, and can lead to improved performance in such applications as trace gas detection, bio-sensing, and molecular characterization~\cite{modern_spectroscopy,Michael2019,Virga2019,Thiel2001FourWaveMA}.

\begin{figure}[htbp!]
	\centering
	\includegraphics[width=\columnwidth]{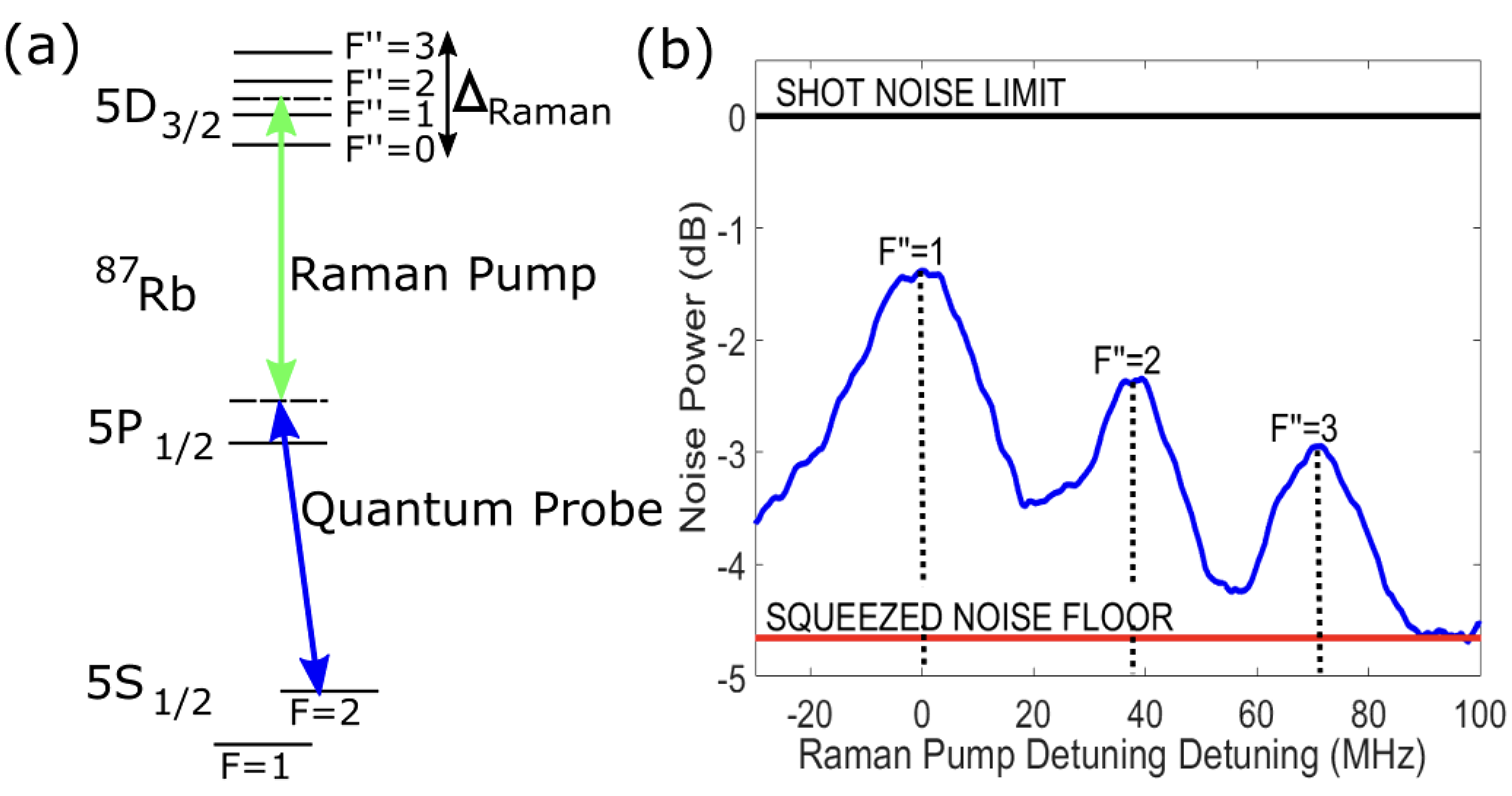}
	\caption{(a) The two-photon transition ($5S_{1/2},F=2\rightarrow 5D_{3/2},F''=1,2,3$) of the probe field coupled to the Raman pump field. (b) Differential intensity noise trace from spectrum analyzer mapping the $5D_{3/2}$ structure as the Raman pump field frequency is tuned through the corresponding two-photon resonances. The Raman pump power is $5$~mW, the cell temperature $65^o$C (atomic density $3.7\cdot 10^{11} \mathrm{cm^{-3}}$).}
	\label{fig:traces}
\end{figure}

Here we present a proof-of-principle demonstration of the proposed method in which we were able to detect the spectrum of the 5D$_{3/2}$ state in $^{87}$Rb via two-photon Raman resonances in the ladder configuration~\cite{zibrov02pra,AkulshinOE09,PrajapatiJOSAB18}, and experimentally show that the use of the two-mode squeezed light allowed us to reduce the minimum required pump power by approximately a factor of $3-5$ for the same signal to noise ratio. 
In these experiments we coupled one of the squeezed twin-beams (probe) with the Raman pumping field in a ladder configuration to probe the $5D_{3/2}$ excited state, level diagram shown in Fig.~\ref{fig:traces}(a). 

The schematic of the experimental setup is shown in Fig.~\ref{fig:setup_raman}, with more details available in Ref.~\cite{Prajapati19OL}. 
Intensity-squeezed twin-beams are generated by mixing the strong pump field (average power 200~mW, beam diameter at the cell 0.45mm) and the weak probe field (average power 40~$\mu$W, beam diameter at the cell 0.34mm) at an angle of 4 mrad in the $^{85}$Rb vapor cell, maintained at the temperature 106$^o$C (labeled ``FWM $^{85}$Rb Cell'' in the diagram). 
After the cell, the pump beam is filtered out using a polarizing beam splitter, and the amplified probe and newly generated conjugate beams display approximately $5$~dB of relative intensity noise reduction below the shot noise limit. 
To optimize the squeezing performance, the pump field is tuned approximately $1.2$~GHz above the $5S_{1/2}~F=2\rightarrow5P_{1/2}~F'$ transition of $^{85}$Rb, resulting in the probe field being detuned by approximately $2.1$GHz above the $5S_{1/2}~F=2\rightarrow5P_{1/2}~F'=1$ transition of $^{87}$Rb, ensuring no noticeable resonant absorption of the probe field by the $^{87}$Rb atoms.

To perform the quantum-enhanced spectroscopy of the $5D_{3/2}$ level, we focused the squeezed twin-beams into a second cell (length $75$~mm, diameter $22$~mm), containing isotopically enriched $^{87}$Rb vapor for the spectroscopy measurements.
The counter-propagating Raman pump optical field (762.1068~nm), produced by a tunable Ti:sapph laser, overlapped at a small angle with only the probe field. When the frequency sum of the probe and Raman pump laser matched the two-photon transition frequency to the 5D$_{3/2}$ excited state, the probe intensity dropped slightly due to the ladder-type Raman absorption. 
To achieve sub-shot noise regime, we record the difference between the intensities of both probe and conjugate twin beams using a balanced photo-detector (BPD), from which the current was read out by a spectrum analyzer (SA).

We can estimate the effect of Raman absorption on the intensity squeezing between the probe and conjugate optical fields by using a simplified model that traces the evolution of the annihilation operators for these fields (denoted $\hat a$ and $\hat b$, correspondingly) first through the FWM cell and then through the spectroscopy cell, starting with a coherent state for the input probe field, and a coherent vacuum for the input conjugate field. First, we  describe the generation of the two-mode quantum field in the FWM process via the following transformation:
\begin{eqnarray}
    \hat a&\rightarrow &\cosh{(s)} \hat a - \sinh{(s)} \hat b^\dagger;\\ 
    \hat b&\rightarrow &\cosh{(s)} \hat b - \sinh{(s)} \hat a^\dagger, \nonumber
\end{eqnarray}
where $s$ is the two-mode squeezing parameter, related to the gain of the four-wave mixing process $G$ as $\cosh^2{(s)}=G$~\cite{lettPRA08}. The additional probe field losses, associated with the two-photon Raman resonance absorption $\alpha(\Delta_R)$, can be modeled using a traditional beam splitter model:
\begin{equation}
    \hat a \rightarrow \sqrt{1-\alpha(\Delta_R)} \hat a + \sqrt{\alpha(\Delta_R)} \hat u_{vac},
\end{equation}
where $\hat u_{vac}$ corresponds to the coherent vacuum. In this case the expected normalized variance of the differential quantum noise can be expressed as following, assuming shot-noise limited input probe field~\cite{jesperse_thesis}:
\begin{eqnarray} \label{eq:raman_noise}
    \frac{\langle \hat N_a - \hat N_b\rangle}{\langle \hat N_a + \hat N_b\rangle}&=&\frac{2G(G-1)\alpha^2-G \alpha+1}{2G-1} \\ &\approx& (G-1)\alpha^2-\frac{\alpha}{2}+\frac{1}{2G-1}. \nonumber
\end{eqnarray}
Away from the Raman absorption resonance ($\alpha_R\approx 0$), the measured noise is determined by the last term that represents  the undisturbed FWM two-mode squeezing. Increasing absorption results in sharp noise increase, dominated by the first term, proportional to the FWM gain $G\gg1$.

In these and following measurements, both probe and conjugate beams travel parallel through the spectroscopy cell to minimize the difference between their optical losses unrelated to the Raman absorption of the probe field. Fig.~\ref{fig:traces}(b) shows the example of the differential intensity noise measurements after the spectroscopy ${}^{87}$Rb cell while the Raman pump laser frequency was tuned across the two-photon Raman transitions. Three absorption lines, corresponding to the transitions to the $5D_{3/2},F''=1,2,3$ states are clearly resolved, even though all the peaks are below the classical shot-noise limit.    

The main advantage of using intensity squeezed twin beams for detection of Raman absorption is that it allows detecting weaker resonances due to improved signal to noise ratio and collecting meaningful absorption spectra for samples with lower effective optical depth. In practice, this translates into reduce requirements for either Raman pump optical power, or the atomic concentration.
 \begin{figure}[htbp!]
	\centering
	\includegraphics[width=\columnwidth]{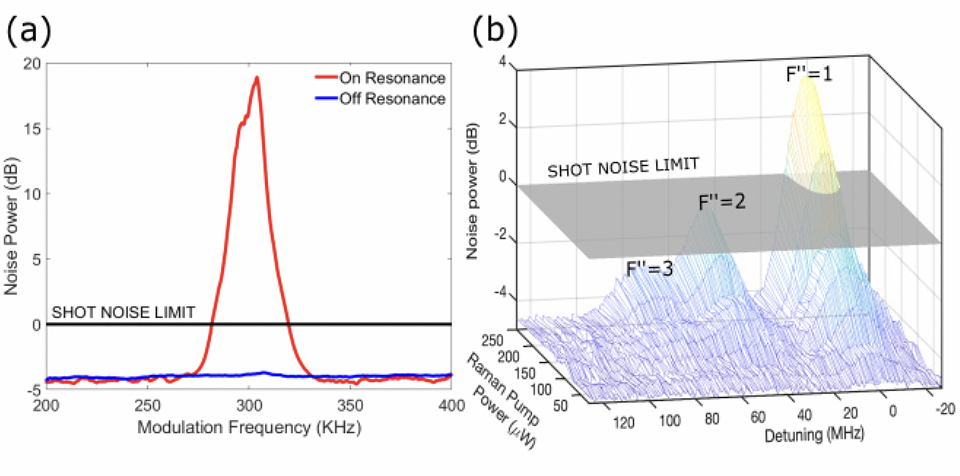}
	\caption{(a) The example of the Raman absorption measurements with amplitude-modulated Raman pump field, showing the signal peak at the Raman pump modulation frequency ($300$~kHz), when the Raman laser is in resonance with the two-photon transition to the $5S_{1/2},F=2\rightarrow5D_{3/2},F''=1$ transition, on top of the reduced noise floor due to balanced detection using intensity squeezed twin beams. (b) Measured absorption spectra mapping the $5D_{3/2}$ structure as the Raman pump field frequency is tuned through the corresponding two-photon resonances for various powers of the coupling laser. The cell temperature $25^o$C (atomic density $9.6\cdot 10^{9} \mathrm{cm^{-3}}$).}
	\label{fig:traces_mod}
\end{figure}

To further boost the measurement sensitivity, we modulated the intensity of the Raman pump at $300$~kHz using an acoustic-optic modulator (AOM), and then monitored any changes in the probe intensity at this modulation frequency. Depending on the proximity to the two-photon resonance and Raman pump power, the peak amplitude will change. The example of recorded noise spectrum on an off the two-photon transition are shown in Fig.~\ref{fig:traces_mod}(a), red trace for resonant conditions and blue trace for off resonant. 
By tracing the peak amplitude, we can obtain the Raman absorption spectrum, similar to that in Fig.~\ref{fig:traces}(b), but with significantly reduced requirement for the Raman pump power compared to the cw case.
Fig.~\ref{fig:traces_mod}(b) shows the measurements of the recorded resonances for different powers of the Raman pump laser. 
It is easy to see that the resonance amplitudes fall quickly with the reduction of the pump power, thus making the measurements with weaker Raman pump more challenging. However, the reduced noise floor for the probe absorption measurements allowed us to clearly resolve the resonance structure down to much lower pump powers compare to the classical detection. For example, for the highest pump power in Fig.~\ref{fig:traces_mod}(b), 250~$\mu$W, only the strongest absorption peak (corresponding to the transition to $F''=1$) was above the shot noise level, while using squeezed light we were able to clearly resolve all three resonances.

\begin{figure}[htbp!]
	\centering
	\includegraphics[width=\columnwidth]{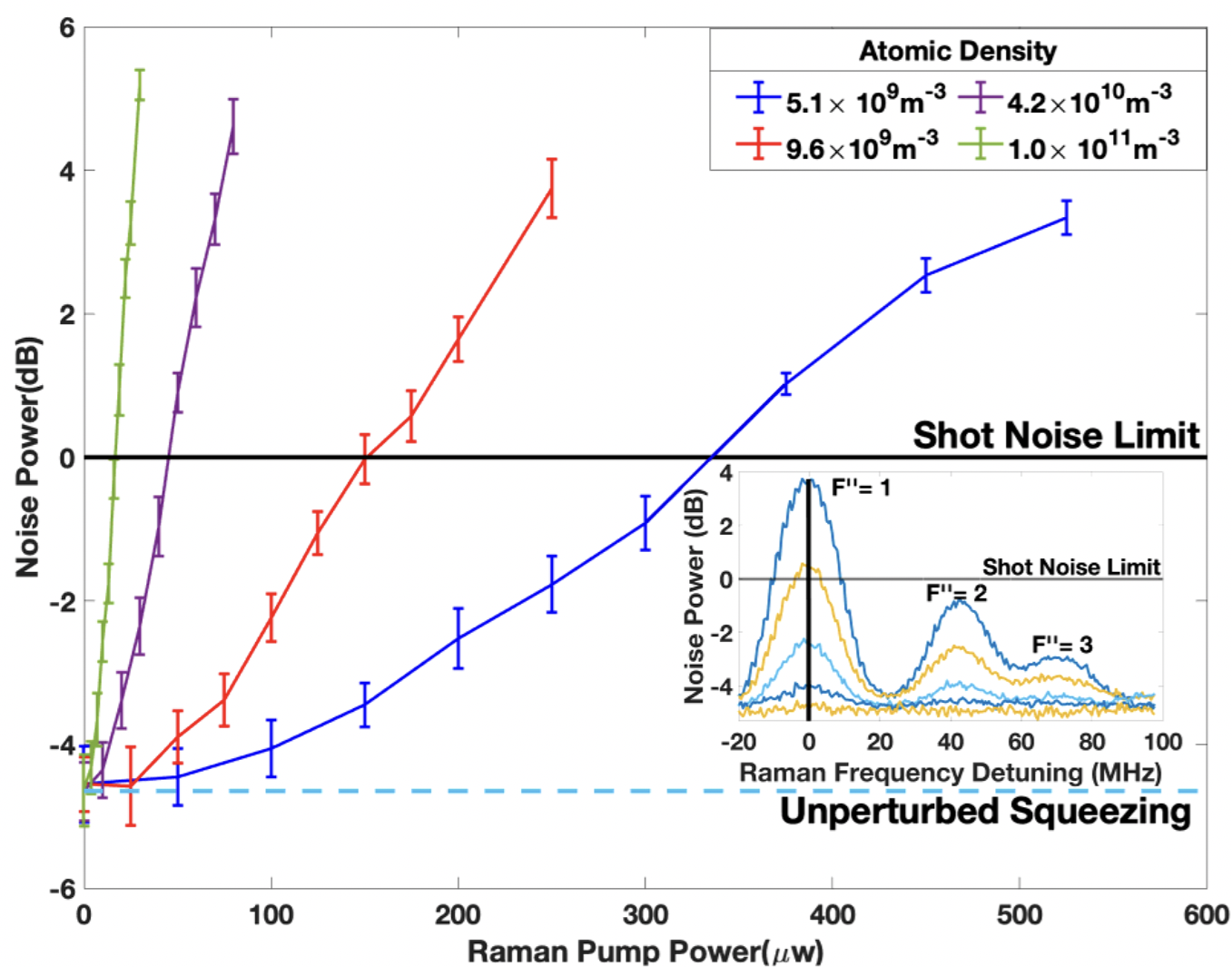}
	\caption{Measured amplitude of the Raman absorption at the $5S_{1/2},F=2 \rightarrow 5D_{3/2},F"=1$ resonance peak as a function of the Raman pump power for different ${}^{87}$Rb cell temperatures. The inset plot, showing the measured spectrum analyzer signal for different Raman pump intensities for the same cell temperature, indicated the location of the measured peak.}
	\label{fig:temp&pow}
\end{figure}

%
To further explore the potential of the quantum enhanced two-photon Raman spectroscopy, we measured the amplitude of the strongest noise peak corresponding to the $5S_{1/2},F=2 \rightarrow 5D_{3/2},F"=1$ optical transition and plotted it as a function of the Raman pump laser power for several different atomic densities of Rb vapor, as shown in  Fig.~\ref{fig:temp&pow}.
The solid line indicate the detection limit set by the classical shot noise, so any data points below this level are only detectable due to the implemented quantum enhanced detection.  
For higher atomic densities ($1.0\cdot 10^{11}\mathrm{cm}^{-3}$, the vapor cell temperature $50^o$C), the use of squeezed beams allowed us to observe the Raman resonance for the probe powers as low as $\approx$4~$\mu$W compare to the classically-limited value of $\approx$20~$\mu$W.
As we decreased the cell temperature, we had to compensate for the reducing number of Rb atoms by increasing the strength of Raman coupling to maintain the similar optical depth. Thus, the power thresholds for both classical and squeezed detection increased for lower atomic densities. Nevertheless, the squeezed optical probe always provided substantial improvement: for example, for the lowest atomic density ($5.1\cdot 10^{9}\mathrm{cm}^{-3}$, the vapor cell temperature $19^o$C), nearly the entire trace was below the classical limit, and the peak was detectable for powers as low as $\approx$50~$\mu$W. 

While the spectroscopy of the Rb $5D_{3/2}$ atomic state is hardly an important accomplishment by its own, this experiment is a proof-of-principle demonstration of the proposed quantum enhanced spectroscopy. By adjusting the frequency of the coupling laser we can realize similar quantum enhancement to characterize the states in the spectral region for which squeezing sources are not available. For example, this method can be immediately applied for sensors based on Rydberg atoms\cite{PhysRevLett.98.113003}, in which the response of the highly excited atomic states is often monitored through the change in optical depth for one of the low-level transitions. Moreover, for many such applications it is desirable to keep the power of such optical probe to a minimum to avoid power broadening, providing additional motivation for relying on squeezing to reduce measurement noise.

Potentially even more far-reaching applications can be achieved for quantum enhanced spectroscopy of molecular samples. In this case one can use an alternative $Lambda$ configuration, shown in Fig.~\ref{fig:setup_raman}~(b), to probe two-photon transitions between the ground state roto-vibrational molecular levels, in a manner similar to stimulated Raman scattering absorption, also known as the inverse Raman effect~\cite{PhysRevLett.13.657,friedrichJCE82}. This would produce a squeezing enhanced Raman spectra for potential use in biology or chemistry~\cite{Nandakumar_2009}, particularly for trace gasses monitoring~\cite{Farrow1894}. 
While the strength of Raman transition for an individual molecule is much lower than for a Rb atom, potentially many more molecules can contribute into total absorption to compensate for weaker transition strength in molecules. 
%
Also, considering that most molecular samples have Raman transitions in the IR spectral region, one may consider alternative sources of squeezing at higher wavelength. For example, squeezed light at telecom wavelength has been generated in fibers, using the optical Kerr effect~\cite{542893}. The highest squeezing has been achieved in PPKTP crystals using frequency down conversion at 1064~nm~\cite{Andersen_2016}. Finally, other alkali metals, in particularly cesium, can be explored for FWM intensity squeezing generation~\cite{PhysRevA.96.043843}.


In conclusion, we proposed a method of using a fixed-frequency source of squeezed light to improve the sensitivity of the spectroscopic measurement via two-photon Raman resonances. In this proof-of-principle demonstration we were able to detect the level structure of the $5D_{3/2}$ state of $^{87}$Rb by means of two-photon Raman absorption resonances with the detection noises 5 dB below the classical shot noise limit using a source of two-mode intensity-squeezed light based on four-wave mixing in ${}^{85}$Rb vapor cell.
Our data demonstrates the clear advantage of using non-classical light as we were able to reduce both pump power and Rb atomic density thresholds and observation of Raman resonances imperceptible for the shot-limited case. 
For example, for the room temperature (Rb density $9.6\cdot 10^9 \mathrm{cm}^{-3}$) the minimum required pump power dropped from $\approx 150~\mu$W at the shot noise limit to $\approx 30~\mu$W. Similarly, for the $\approx 50~\mu$W pump power we were able to resolve the resonances at Rb density as low as $5.1\cdot 10^9 \mathrm{cm}^{-3}$, a factor of 7 reduction from the estimated $\approx 3.45\cdot 10^{10} \mathrm{cm}^{-3}$ for a shot-noise limited measurements. Thus, we believe that such quantum enhanced spectroscopy will find many applications, from precise measurement of the energy level structure of highly excited atoms, such as Rydberg atoms, to enhanced detection of molecules for trace gas detection and biological sensing.
\\
\textbf{Funding}: Air Force Office of Scientific Research  (FA9550-19-1-0066).  \\
\textbf{Acknowledgments}: We would like to thank Eugeniy Mikhailov and Nathan Kidwell for useful discussions and help with the experiment.\\
\textbf{Disclosures}: The authors declare no conflicts of interest.

\bibliography{bibliography1}

\end{document}